\documentclass[prd,a4paper,showpacs,byrevtex,preprint]{revtex4-1}
\usepackage{graphicx}
\usepackage{dcolumn}
\usepackage{amsmath}
\usepackage{array}
\usepackage{bm}
\usepackage{amssymb}
\usepackage{amsfonts}
\begin{document}

\title{Comments on Yang-Mills thermodynamics, the Hagedorn spectrum and the gluon gas}

\author{Fabien \surname{Buisseret}}
\email[E-mail: ]{fabien.buisseret@umons.ac.be}
\affiliation{Service de Physique Nucl\'{e}aire et Subnucl\'{e}aire,
Universit\'{e} de Mons--UMONS,
Acad\'{e}mie universitaire Wallonie-Bruxelles,
Place du Parc 20, B-7000 Mons, Belgium; \\
Haute Ecole Louvain en Hainaut (HELHa), Chauss\'ee de Binche 159, B-7000 Mons, Belgium}

\author{Gwendolyn \surname{Lacroix}}
\email[E-mail: ]{gwendolyn.lacroix@umons.ac.be}
\affiliation{Service de Physique Nucl\'{e}aire et Subnucl\'{e}aire,
Universit\'{e} de Mons--UMONS,
Acad\'{e}mie universitaire Wallonie-Bruxelles,
Place du Parc 20, B-7000 Mons, Belgium}

\begin{abstract}
We discuss the dependence of pure Yang-Mills equation of state on the choice of gauge algebra. In the confined phase, we generalize to an arbitrary simple gauge algebra Meyer's proposal of modelling the Yang-Mills matter by an ideal glueball gas in which the high-lying glueball spectrum is approximated by a Hagedorn spectrum of closed-bosonic-string type. Such a formalism is undefined above the Hagedorn temperature, corresponding to the phase transition toward a deconfined state of matter in which gluons are the relevant degrees of freedom. Under the assumption that the renormalization scale of the running coupling is gauge-algebra independent, we discuss about how the behavior of thermodynamical quantities such as the trace anomaly should depend on the gauge algebra in both the confined and deconfined phase. The obtained results compare favourably with recent and accurate lattice data in the $\mathfrak{su}(3)$ case and support the idea that the more the gauge algebra has generators, the more the phase transition is of first-order type. 
\end{abstract}


\maketitle

\section{Introduction}
The existence of a critical temperature, $T_c$, in QCD, is of particular phenomenological interest since it  signals a transition from a confined phase of hadronic matter to a deconfined one. When $T<T_c$, a successful effective description of QCD is the hadron resonance gas model, in which the hadronic matter is seen as an ideal gas of hadrons. It compares well with current lattice data when the meson and baryon resonances below 2.5~GeV are included~\cite{borsanyi2010}. A problem is that experimental information about resonances above 3~GeV is still lacking. To describe the high-lying hadronic spectrum, Hagedorn~\cite{hage65} proposed a model in which the number of hadrons with mass $m$ is found to increase as $\rho(m)\propto m^a \, {\rm e}^{m/T_h}$ ($a$ is real): the so-called Hagedorn spectrum. Thermodynamical quantities, computed using hadronic degrees of freedom, are then undefined for $T>T_h$. Other degrees of freedom are then needed at higher temperatures, so it is tempting to guess that $T_h\approx T_c$, the new degrees of freedom being deconfined quarks and gluons. 

Although the current lattice studies agree on a value of $T_c$ in the range $(150-200)$~MeV when $2+1$ light quark flavours are present~\cite{borsanyi2010,tcd}, there is currently no consensus concerning the value of $T_h$. Indeed, to reach values of $T_h$ as low as 200~GeV demands an ad hoc modification of $\rho(m)$: By introducting an extra parameter $m_0$ and setting $\rho(m)\propto (m^2+m^2_0)^{a/2} \, {\rm e}^{m/T_h}$, one can reach values of $T_h$ in the range $(160-174)$ MeV, that agree with lattice computations, see \textit{e.g.}~\cite{hage68,cley}. However, by taking the original form $m_0=0$, one rather ends up with values of $T_h$ around $(300-360)$~MeV, see \cite{cudell0,bronio}. Moreover, it has been observed in some pure gauge lattice simulations with the gauge algebra $\mathfrak{su}(N)$ that $T_c\lesssim T_h$~\cite{TcTh0,TcTh} as intuitively expected. It has to be said that the value of $T_h$ and its relation to $T_c$ are still a matter of debate. 

Open strings as well as closed strings naturally lead to a Hagedorn spectrum, see \textit{e.g.} \cite{zwie}. Modelling mesons as open strings is a way to make appear a Hagedorn spectrum in QCD~\cite{cudell}. The question of showing that a Hagedorn spectrum arises from QCD itself is still open but, under reasonable technical assumptions, it has recently been found in the large-$N$ limit of QCD~\cite{cohen} (glueballs and mesons have a zero width in this limit). In the pure gauge sector, the $\mathfrak{su}(3)$ equation of state computed on the lattice has been shown to be compatible with a glueball gas model in which the high-lying spectrum is modelled by a gas of closed bosonic strings~\cite{meyer}.  

Besides QCD, pure Yang-Mills (YM) thermodynamics is challenging too, in particular because it can be formulated for any gauge algebra. A clearly relevant case is the one of $\mathfrak{su}(N)$-type gauge algebras, linked to the large-$N$ limit of QCD. Moreover, a change of gauge algebra may lead to various checks of the hypothesis underlying any approach describing $\mathfrak{su}(3)$ YM theory. To illustrate this, let us recall the pioneering work~\cite{sve}, suggesting that the phase transition of YM theory with gauge algebra $\mathfrak{g}$ is driven by a spontaneous breaking of a global symmetry related to the center of $\mathfrak{g}$. Effective $Z_3$-symmetric models are indeed able to describe the first-order phase transition of $\mathfrak{su}(3)$ YM thermodynamics~\cite{Z3}. However, a similar phase transition has also been observed in lattice simulations of G$_2$ YM theory~\cite{G2} even though the center of G$_2$ is trivial, meaning that the breaking of center symmetry is not the only mechanism responsible for deconfinement. For example, it is argued in~\cite{diakonov} that the YM phase transition for any gauge group is rather driven by dyons contributions. In this case, still under active investigation, studying different gauge algebras helps to better understand the general mechanisms of (de)confinement in YM theory. For completeness, we mention that the structure of the gluon propagator at low momentum as well as the Dyson-Schwinger equations in scalar-Yang-Mills systems have recently started to be studied for generic gauge algebra~\cite{maas,maas2}.

The main goal of the present work is to give predictions for the equation of state of YM theory with an arbitrary simple gauge algebra. This topic has, to our knowledge, never been investigated before and will be studied within two well-established different frameworks: A glueball gas with a high-lying Hagedorn spectrum in the confined phase (Sec.~\ref{conf}) and a gluon gas above the critical one (Sec.~\ref{deconf}). Some phenomenological consequences of the obtained results will then be discussed in Sec. \ref{conclu}. More specifically, our results apply to the following gauge algebras : A$_{r\geq 1}$ related to $\mathfrak{su}$ algebras, B$_{r\geq 3}$ and D$_{r\geq 4}$ related to $\mathfrak{so}$ algebras, C$_{r\geq 2}$ related to  $\mathfrak{sp}$ algebras, and the exceptional algebras E$_{6}$, E$_7$, F$_4$ and G$_2$. The case of E$_8$ is beyond the scope of the present paper as it will be explained below.

\section{Glueball gas and the Hagedorn spectrum}\label{conf}
\subsection{The model}
In the confined phase, glueballs, \textit{i.e.} colour singlet bound states of pure YM theory, are the relevant degrees of freedom of YM matter. Hence it can be modelled in a first approximation by an ideal gas of glueballs, assuming that the residual interactions between these colour singlet states are weak enough to be neglected~\cite{dashen}. Note that the glueball gas picture emerges from a strong coupling expansion in the case of large-$N$ $\mathfrak{su}(N)$ YM theory~\cite{langelage10}, where glueballs are exactly noninteracting since their scattering amplitude scales as $1/N^2$~\cite{witten}. The glueball gas picture implies that, for example, the total pressure should be given by $\sum_{J^{PC}}p_0(2J+1,T,M_{J^{PC}})$, where the sum runs on all the glueball states of the YM theory with a given gauge algebra, and where
\begin{equation}
p_0(d,T,M)=\frac{d}{2\pi^2}M^2T^2\sum_{j=1}^\infty\frac{1}{j^2}K_2(j\, M/T)
\end{equation}
is the pressure associated with a single bosonic species with mass $M$ and $d$ degrees of freedom.  

Performing the sum $\sum_{J^{PC}}$ demands the explicit knowledge of all the glueball states, not only the lowest-lying ones that can be known from lattice computations or from effective approaches. To face this problem, it has been proposed in \cite{meyer} to express the total pressure of $\mathfrak{su}(3)$ YM theory as
\begin{equation}\label{preh}
p= \hspace{-0.3cm}\sum_{M_{J^{PC}}<2M_{0^{++}}}\hspace{-0.65cm} p_0(2J+1,T,M_{J^{PC}})+\int^\infty_{2M_{0^{++}}}\hspace{-0.5cm}dM\ p_0(\rho(M),T, M),
\end{equation}
where the high-lying glueball spectrum (above the two-glueball threshold $2M_{0^{++}}$) is approximated by a closed-string Hagedorn density of states reading, in 4 dimensions~\cite{zwie,meyer},
\begin{equation}\label{rhoh}
\rho(M)=\frac{(2\pi)^3}{27 T_h}\left( \frac{T_h}{M} \right)^4{\rm e}^{M/T_h}.
\end{equation}
The idea of modelling glueballs as closed fundamental strings was actually already present in the celebrated Isgur and Paton's flux-tube model, inspired from the Hamiltonian formulation of lattice QCD at strong coupling~\cite{isgur}. Moreover, it has also been shown within a constituent picture that, in the $\mathfrak{su}(3)$ case, a many-gluon state (typically more than three gluons in a Fock-space expansion) tends to form a closed gluon chain~\cite{gluphen}. 

In Eq.~(\ref{rhoh}), $T_h$ is the Hagedorn temperature, which reads in this case
\begin{equation}
T^2_h=\frac{3}{2\pi}\sigma^{(f)},
\end{equation}
where $\sigma^{(f)}$ is the fundamental string tension, here defined as the slope of the static energy between two sources in the fundamental representation of a given gauge algebra. The Casimir scaling of the string tension, which is an analytic prediction from the strong coupling expansion of the Wilson loop, says that the string tension is given by~\cite{casi,casi2} 
\begin{equation}
\sigma^{(r)}=C_2^{(r)}\, \Theta  ,
\end{equation}
where the colour sources are in a given representation $r$ of the gauge algebra, and where $\Theta$ reads, in a lattice formulation of the theory~\cite{casi}
\begin{equation}
\Theta=\frac{g^2(a\Lambda)}{2a}.
\end{equation}
$a$ is the lattice size and $g(a\Lambda)$ is the running coupling with the renormalization scale $\Lambda$. Following well-known two-loop calculations, one can extract the explicit gauge-algebra dependence in the running coupling as follows: $g^2(a\Lambda)=\lambda(a\Lambda)/C_2^{(ad\hspace{0.1pt}j)}$~\cite{g2r}, where $\lambda$ is nothing else than the 't Hooft coupling when the gauge algebra is $\mathfrak{su}(N)$. One can finally define 
\begin{equation}
\sigma^{(r)}=\frac{C_2^{(r)}}{C_2^{(ad\hspace{0.1pt}j)}}\, \sigma_0,
\end{equation}
where $\sigma_0$, that can be interpreted as the adjoint string tension, does not depend explicitly on the gauge algebra. However, an implicit dependence in the renormalization scale $\Lambda$ may be present. Throughout this work we consider a gauge-algebra independent value for $\Lambda$. 

The structure of the low-lying glueball spectrum for an arbitrary simple gauge algebra has been discussed in detail in~\cite{buiss11} within a constituent picture, although the results which are useful for our purpose could be recovered in a more model-independent way by studying \textit{e.g.} the structure of glueball-generating field-strength correlators. Let us recall those results:
\begin{itemize}
\item The lightest glueballs are the scalar, pseudoscalar and tensor ones, whose masses are ordered as $M_{0^{++}}<M_{2^{++}}$, $M_{0^{-+}}$ in agreement with lattice results in the $\mathfrak{su}(N)$ case~\cite{glueb1,luciN}. Those states are found to be lighter than $2\, M_{0^{++}}$ in these last works. Note that it has been proved in ~\cite{west} that the $0^{++}$ glueball is always the lightest one in YM theory. 
\item At masses typically around (3/2)$M_{0^{++}}$, states that can be seen as mainly three-gluon ones in a Fock-space expansion appear: They can have $C=+$ for any gauge algebra, but $C=-$ for A$_{r\geq2}$ ($\mathfrak{su}(N\geq 3)$) only. In this last case, the $1^{+-}$ glueball is still lighter than $2\, M_{0^{++}}$~\cite{glueb1,luciN}.
\item Higher-lying states (containing more than three gluons in a Fock space expansion) obviously exist, but their exhaustive study cannot be performed explicitly, eventually justifying the use of a Hagedorn spectrum. An important remark has nevertheless to be done: If all the representations of a given gauge algebra are real, the gluonic field $A_\mu $ is its own charge-conjugate, eventually forbidding $C=-$ glueball states. This happens for the algebras A$_1$, B$_{r\geq2}$, C$_r$, D$_{{\rm even}-r\geq4}$, E$_7$, E$_8$, F$_4$, and G$_2$. 
\end{itemize}

It is worth noticing that a closed-string picture for high-lying glueballs is not only a consequence of Isgur and Paton's flux-tube-like approaches but may also be compatible with constituent approaches such as the one used in~\cite{buiss11}: An excited closed string is then alternatively viewed as a closed chain of quasigluons where the quasigluons are linked by fundamental strings. From a string theory point of view, the Nambu-Goto string can be coherently quantized within both pictures using \textit{e.g.} the Gupta-Bleuler method~\cite{gershun10}. Moreover, since $ad\hspace{0.1pt}j\in f\otimes f$ or $f\otimes\bar f$, with $f$ ($\bar f$) the fundamental (conjugate) representation for any simple gauge algebra, a gluon can always generate two fundamental strings, with $\sigma^{(f)}=\sigma^{(\bar f)}$ in virtue of the Casimir scaling, instead of one adjoint string. In the case of E$_8$, the lowest-dimensional representation, that we have called fundamental before, is the adjoint one, so the closed-string picture seems less justified by comparison to a constituent picture. We therefore prefer not to investigate further the case of E$_8$ in the following.

\subsection{Linking $T_h$ to $T_c$}

As a first step, the link between $T_h$ and $T_c$ has to be fixed. A straightforward way to do it is to briefly recall Meyer's results in the pure gauge $\mathfrak{su}(3)$ case~\cite{meyer}, where the lattice entropy density $s=\partial_T p$ computed below $T_c$ has been fitted by using the present model. It appears that the best agreement is reached for $T_h/T_c=1.069(5)$. Finding $T_h>T_c$ is actually an indication that a metastable, superheated, hadronic phase of matter exists at temperatures between $T_c$ and $T_h$; this phase has actually been studied on the lattice in \cite{TcTh}, where, for example, $T_h/T_c=1.116(9)$ has been found for the gauge algebra $\mathfrak{su}(12)$, and discussed within the framework of an open-string model in \cite{cudell}. 

As seen from the above discussion, an accurate determination of the ratio $T_h/T_c$ is of great phenomenological interest. However, such a study is not the main purpose of the present paper, where we aim at giving reliable predictions for the equation of state of YM theory with an arbitrary gauge algebra. As observed in~\cite{meyer}, typical values $T_h\approx T_c$ give very good results in fitting the lattice data. Setting $T_c=T_h$, as we will do in the rest of this work, means that the deconfinement temperature may be identified with the maximal allowed temperature for the confined hadronic phase. This assumption has two advantages. First, it will reproduce accurately the latest $\mathfrak{su}(3)$ lattice data of~\cite{borsa} (see next section), and it is not in strong disagreement with current $\mathfrak{su}(N)$ results, where $T_h/T_c$ is at most around $10 \%$~\cite{TcTh0,TcTh}. Second, it is applicable to any gauge algebra without having to guess a value for $T_h/T_c$, that cannot be fitted on lattice results since no equation of state is available for gauge algebras different than $\mathfrak{su}(N)$ so far. The drawback of this choice is that it forbids any discussion about a superheated hadronic phase in generic YM theories. Such a refinement of the model will rather be the topic of a separate study. 

For completeness, we notice that the somewhat surprising value $T_h=2.8\, T_c\gg T_c$ has been found in~\cite{Megias} by using a Hagedorn picture too. The difference with our approach comes from the fact that, in ~\cite{Megias}, $T_h$ is fitted by assuming that the low-lying glueballs currently known from lattice simulations should exhibit a Hagedorn-type spectrum. On the contrary, we think here that the Hagedorn-like behavior only appears in the high-lying sector, that mostly concerns the glueballs that are not known so far by lattice calculations, see Eq.~(\ref{preh}).

\subsection{Numerical results}
According to standard $\mathfrak{su}(3)$ studies, it is relevant to set $\sigma_0\approx(9/4)\ 0.2$~GeV$^2$, leading to $T_h=$309~MeV. The masses of the lightest glueballs are proportional to $\sqrt{\sigma_0}$~\cite{buiss11}, so they can be thought as constant with respect to a change of gauge algebra in our approach. Consequently, the sum $\sum_{M_{J^{PC}}<2M_{0^{++}}}$ should run on all the states below $3.46$~GeV found in the $\mathfrak{su}(3)$ lattice work~\cite{glueb1}. There is an exception however: The $1^{+-}$ glueball, whose mass is below the two-glueball threshold, only exists when the gauge algebra is A$_{r\geq2}$~\cite{buiss11}; hence its contribution will be omitted in the other cases. Concerning the Hagedorn spectrum, it is worth recalling that the density~(\ref{rhoh}) is able to reproduce the $\mathfrak{su}(3)$ lattice equation of state with $T_c\approx T_h$~\cite{meyer}. But $\rho(M)$ accounts for both the $C=+$ and $C=-$ glueballs. When the gauge algebra has only real representations, the $C=-$ sector is absent as said before. So in such cases, the substitution $\rho(M)\rightarrow\rho(M)/2$ will be done. The validity of this prescription has been explicitly checked in~\cite{case} by computing the equation of state of $2+1$-dimensional YM theory below $T_c$ with $\mathfrak{su}(N)$ gauge algebras: $\rho(M)$ correctly describes the data for $\mathfrak{su}(3-6)$, while $\rho(M)/2$ must be used for $\mathfrak{su}(2)$ in order to compensate for the absence of $C=-$ states in the theory.       
  
\begin{figure}[t]
\includegraphics*[width=10.0cm]{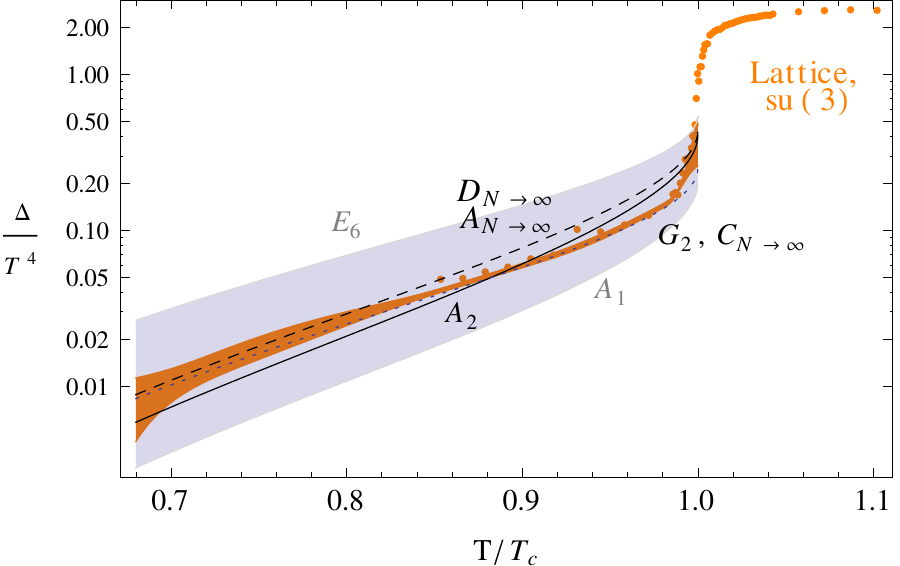}
\caption{(Color online) Trace anomaly below $T_c$, computed using Eqs.~(\ref{preh}) and (\ref{taoh}) with $T_h=T_c$ and $\sigma_0=(9/4)0.2$~GeV$^2$, for the gauge algebras A$_2$ (solid line), A$_{N\rightarrow\infty}$ and D$_{N\rightarrow\infty}$ (dashed line), G$_2$ and C$_{N\rightarrow\infty}$ (dotted line). All the possible cases are located within the grey area, whose upper and lower borders are E$_6$ and A$_1$ respectively. $\mathfrak{su}(3)$ lattice data from \cite{borsa} are plotted for comparison (orange points and area). The orange points correspond to $N_t=8$ data.}
\label{fig1}
\end{figure}

We are now in position of explicitly computing the pressure~(\ref{preh}) for any gauge algebra, E$_8$ excepted. We actually compute from $p$ the trace anomaly, using
\begin{equation}\label{taoh}
\Delta=T^5\partial_T \left(\frac{p}{T^4}\right),
\end{equation}
so that our results can be compared to the recent and accurate $\mathfrak{su}(3)$ lattice data of~\cite{borsa}, displayed in Fig.~\ref{fig1}. 

As a first check, we can see that the proposed model compares well with the $\mathfrak{su}(3)$ lattice data of \cite{borsa}. In a first approximation, the choice $T_c=T_h$ thus gives good results. A generic feature of $p$ and $\Delta$ is that they are finite in $T_h$, and mostly located below the E$_6$ and A$_1$ cases at any $T$. This finiteness is due to the $M^{-4}$ factor in (\ref{rhoh})~\cite{frautschi}, which is a consequence of the closed-string picture used here. Note that this finiteness is present in $2+1$ dimensions too~\cite{case}. An interesting feature is that the large-$N$ limits of the A$_N$ and D$_N$ (when $N$ is odd) cases are equivalent, in agreement with the large-$N$ orbifold equivalence between $\mathfrak{su}(N)$ and $\mathfrak{so}(2N)$ YM theories, see \textit{e.g.}~\cite{cher}. The large-$N$ limit of the C$_N$ ($\mathfrak{sp}(N)$) case is however inequivalent to the A$_N$ one, but equal to the G$_2$ case. The observed significant numerical differences between the gauge algebras are moreover relevant from a physical point of view since they come from changes in the structure of the glueball spectrum, mainly at the level of the allowed quantum numbers. 

It is worth mentioning that an alternative to the Hagedorn spectrum has been proposed in \cite{strange}, \textit{i.e.} to consider that a Hagedorn spectrum is not present but that the glueball masses actually decrease near the critical temperature. This scenario can also lead to an agreement with the data of \cite{borsa} as checked by the authors of this last work. Only the lightest glueballs ($0^{\pm+}$ and $2^{++}$) will then give relevant thermodynamical contributions for any gauge algebra, and the corresponding equation of state might depend even less on the gauge algebra than within the Hagedorn picture. However, checking the dependence on $T$ of the glueball masses for different gauge algebras would demand detailed lattice computations or effective models that are currently unavailable, thus this topic is out of the scope of the present paper. 

\section{Gluon gas and the deconfined phase}\label{deconf}

As already mentioned, the Hagedorn temperature can be interpreted as a limiting temperature above which confined matter ceases to exist. In the deconfined phase, the relevant degrees of freedom are expected to be the ${\rm dim}(ad\hspace{0.1pt}j)$ gluons of the considered YM theory. When the temperature tends toward infinity, the Stefan-Boltzmann limit should thus be reached, that is \textit{e.g.} the pressure
\begin{equation}
p_{SB}={\rm dim}(ad\hspace{0.1pt}j)\, \frac{\pi^2}{45}T^4,
\end{equation} 
corresponding to the pressure an ideal gas of massless transverse gluons with ${\rm dim}(ad\hspace{0.1pt}j)$ colour degrees of freedom in $3+1$ dimensions. Corrections to this ideal-gas picture are nevertheless worth to study since it is known from $\mathfrak{su}(3)$ lattice simulation that one has to reach temperatures of about $(10^7-10^8)$ $T_c$ to get pressures compatible with the Stefan-Boltzmann limit up to the error bars~\cite{highT}. 

The YM pressure (as well as cases with $N_f\neq 0$) can be systematically computed by performing expansions in the coupling constant $g$; terms of order $g^6 \ln(1/g)$~\cite{Laine} and parts of the full $g^6$ terms~\cite{Laine2} are known so far. Hard-thermal-loop (HTL) resummation techniques also allow for a determination of YM pressure; results at next-to-next-to leading order (NNLO) are nowadays available~\cite{HTL}. Recalling the scaling $ g^2\propto 1/C_2^{(ad\hspace{0.1pt}j)}$, the observation of the formulas obtained in~\cite{Laine,HTL} lead to the conclusion that the pressure behaves schematically as
\begin{equation}\label{phT}
\frac{p}{p_{SB}}\equiv 1- \phi(\Lambda ,T)
\end{equation}
where $\Lambda$ is a renormalization scale, that we assume to be gauge-independent as before, and where $\phi$ is a positive function that decreases when $T$ increases so that the SB limit is asymptotically reached. Once the a priori unknown parameters are fitted, both the $O(g^6 \ln(1/g))$ and the NNLO HTL formulae compares very well with the latest $\mathfrak{su}(3)$ lattice data of \cite{borsa}, the best agreement being reached with the $O(g^6 \ln(1/g))$ formula. In particular, the trace anomaly $\Delta$, given by
\begin{equation}\label{delta}
\frac{\Delta}{p_{SB}}=T\, \partial_T\left(\frac{p}{p_{SB}}\right),
\end{equation}
is accurately reproduced above 10 $T_c$ (plots range from 1 to 100 $T_c$ in \cite{borsa}). 

One is straightforwardly led to the conclusion that the pressure (\ref{phT}) is gauge-algebra independent; hence the high-temperature regime of YM thermodynamics should not depend on the considered gauge algebra once the equation of state is normalized to ${\rm dim}(ad\hspace{0.1pt}j)$. For example, the normalized trace anomaly (\ref{delta}) should be gauge-algebra independent. This feature has already been checked on the lattice in the $\mathfrak{su}(N)$ case, where it appears that the pure YM equation of state normalized to $(N^2-1)$ is indeed universal above $T_c$ up to the error bars~\cite{case,panero,gupta}. 

Just above $T_c$, where HTL or perturbative methods cannot give reliable information so far because of convergence problems, gluon-gluon interactions are expected to be quite strong although not confining. One would then speak of strongly coupled YM plasma. Those interactions, typically of one-gluon-exchange form, should be proportional to the color factor $(C_2^{(r)}-2C_2^{(ad\hspace{0.1pt}j)})g^2/2$, where $r$ is the color representation of the gluon pair. The universality of static colour interactions, once normalized to this last colour factor, has been checked on the lattice in the $\mathfrak{su}(3)$ case~\cite{scalT}. For any algebra, one has $ad\hspace{0.1pt}j\otimes ad\hspace{0.1pt}j=\bullet\oplus ad\hspace{0.1pt}j\oplus \dots$. The singlet ($\bullet$) and adjoint channels will lead to attractive interactions that should not depend on the gauge algebra since $g^2\propto1/C_2^{(ad\hspace{0.1pt}j)}$. Other representations appearing in this tensor product will have larger values of $C_2^{(r)}$ and will lead to either weakly attractive, vanishing, or repulsive interactions that may eventually be gauge-algebra dependent. The interesting point is that the most attractive channel is that of a colour-singlet gluon pair, which should not depend on the considered gauge algebra and which is eventually able to form glueballs. So the glueball formation (or not) above deconfinement might well be a universal feature of YM theory; arguments favoring the existence of glueballs beyond $T_c$ have been given for example in \cite{brau}. We mention finally that, in the case of $\mathfrak{su}(N)$ gauge algebras, each channel of the tensor product $ad\hspace{0.1pt}j\otimes ad\hspace{0.1pt}j$ has been explicitly computed in \cite{buiss10}. Two channels lead to weak $N$-dependent interactions (with a $1/N$ colour factor) that may lead to some subleading $N$-dependent corrections. 

\section{Summary and discussion}\label{conclu}

To summarize, we have discussed two pictures of YM matter that allow to compute its thermodynamical properties for any gauge algebra. In the confined phase, the relevant degrees of freedom are glueballs, whose low-lying states can be separately described, while the high-lying states are modelled by a closed bosonic string Hagedorn spectrum. Such a spectrum exhibits a Hagedorn temperature, above which hadronic matter ceases to exist: The partition function of a glueball gas with Hagedorn spectrum is not defined above $T_h$, suggesting a phase transition to a deconfined regime. In the deconfined phase, YM thermodynamics should be the one of an interacting gluon gas. 

In the confined phase, the present model compares favorably with the recent pure gauge $\mathfrak{su}(3)$ lattice data of~\cite{borsa} with a standard value $(9/4)\, 0.2$~GeV$^2$ for the adjoint string tension and the assumption $T_c=T_h$. This does not excludes that a better fit can be found with $T_h\gtrsim T_c$ as in~\cite{meyer}, or that the value $T_c=T_h$ is an artifact due to the simplicity of the model, especially near $T_c$. But, the success of equating $T_c$ and $T_h$ also suggests that the temperature range in which a metastable hadronic phase exists is quite small with the gauge algebra $\mathfrak{su}(3)$. Keeping the relation $T_c=T_h$ as well as the value of the adjoint string tension unchanged, predictions for the equation of state of YM theory with arbitrary gauge algebras have been given; it can be hoped that future lattice simulations will be able to confirm them (or not), at least is some cases of current interest like YM theory with $G_2$ gauge algebra. 
  
It is worth saying that identifying the critical temperature to the Hagedorn temperature leads to the possibility of estimating the gauge-algebra dependence of $T_c$. A relevant example is that, in the case of $\mathfrak{su}(N)$ gauge algebras, we are led to the prediction that $T_c[\mathfrak{su}(2)]/T_c[\mathfrak{su}(\infty)]=\sqrt{3}/2=0.866$, which can be favourably compared to the Polyakov-loop based approach~\cite{braun} finding the value 0.898 for this last ratio. For a $\mathfrak{sp}(2)$ gauge algebra, we find $T_c[\mathfrak{sp}(2)]/T_c[\mathfrak{su}(\infty)]=\sqrt{5/6}=0.913$ while a comparable ratio of $0.969$ is found in \cite{braun}.

Our framework implies that the thermodynamical observables are of $O((d-1)\times{\rm dim}(ad\hspace{0.1pt}j))$ above $T_c$ for a Yang-Mills theory in $d+1$ dimensions and a gauge algebra having ${\rm dim}(ad\hspace{0.1pt}j)$ generators. Consequently, these observables should of $O(1)$ when both $C=+$ and $-$ glueballs are present, \textit{i.e.} for A$_{r\geq2}$, D$_{{\rm odd}-r\geq5}$, and E$_6$, and of $O(1/2)$ in the other cases. The pressure ratio
\begin{equation}
\delta=\lim_{\eta\rightarrow 0}\frac{p(T_c+\eta)}{p(T_c-\eta)},
\end{equation}
where $\eta$ is positive, is then generally of order $2(d-1){\rm dim}(ad\hspace{0.1pt}j)$, but of order $(d-1){\rm dim}(ad\hspace{0.1pt}j)$ for A$_{r\geq2}$, D$_{{\rm odd}-r\geq5}$, and E$_6$\footnote{Note that $T^2_h=3\sigma^{(f)}/(\pi(d-1))$ in $d+1$ dimensions, but this $d$-dependence does not affect the order of magnitude of $\delta$.}. More explicitly, $\delta=16$ for $\mathfrak{su}(3)$ in $3+1$ dimensions, a case for which the phase transition is known to be weakly first order. Some cases can be mentioned for which $\delta\ll 16$: $\mathfrak{su}(2)$ in $3+1$ dimensions and $\mathfrak{su}(2,3)$ in $2+1$ dimensions. It is tempting to say that such small gaps should lead to a second order phase transition. Although the argument seems quite naive, this is indeed the case: It is known from lattice simulations that the phase transition is of second order in those cases~\cite{case}. Moreover, $\delta=15\approx16$ for $\mathfrak{su}(4)$ in $2+1$ dimensions, presumably leading to a (very) weakly first-order phase transition, as observed in~\cite{case}. Moreover, $\delta\gg 16$ for $\mathfrak{su}(N>3)$ in $3+1$ dimensions, corresponding to a phase trantision more and more of first-order type for $\mathfrak{su}(N)$ when $N$ increases, in agreement with previous lattice results~\cite{TcTh0}. It seems thus that our picture eventually leads to criterion allowing to guess the strength of the deconfining phase transition in YM theories. Note that, according to this criterion, any gauge algebra for Yang-Mills theory in $2+1$ and $3+1$ dimensions should lead to a first-order phase transition, $\mathfrak{su}(2)$ ($\mathfrak{su}(2,3)$) in $3+1$ $(2+1)$ dimensions excepted.  

Finally, these results can be linked to an already proposed argument, saying that the mismatch of the number of degrees of freedom above and below the phase transition is responsible for the weakly or strongly first-order character of the deconfinement phase transition~\cite{G2,holl,braun}. Here we reach the same conclusion up to a little difference: The number of glueballs, \textit{i.e.} the relevant degrees of freedom in the confined phase, is formally infinite but leads to thermodynamical contributions that do not directly depend on ${\rm dim}(ad\hspace{0.1pt}j)$, while the gluons, that control the thermodynamics in the deconfined phase, are finite in number but lead to thermodynamical contribtutions proportional to ${\rm dim}(ad\hspace{0.1pt}j)$.

\section*{Acknowledgements}
FB thanks the F.R.S.-FNRS for financial support. GL thanks the UMons for financial support. We thank Sz. Borsanyi and G. Endrodi for having provided us the data of \cite{borsa}, and M. Laine and M. Panero for interesting remarks about the manuscript.


\begin{thebibliography}{99}
\bibitem{borsanyi2010} S.~Borsanyi \textit{et al.}, JHEP {\bf 1011}, 077 (2010) [arXiv:1007.2580].
\bibitem{hage65} R.~Hagedorn, Nuovo Cim.\ Suppl.\  {\bf 3}, 147 (1965).
\bibitem{tcd} M. Cheng {\it et al.}, Phys.\ Rev.\  D {\bf 74}, 054507 (2006) [hep-lat/0608013]; Y. Aoki {\it et al.}, JHEP {\bf 0906}, 088 (2009) [arXiv:0903.4155].
\bibitem{hage68} R. Hagedorn, Nuovo Cim. A \textbf{56}, 1027 (1968).
\bibitem{cley} J.~Cleymans and D.~Worku, Mod. Phys. Lett. A \textbf{26}, 1197 (2011) [arXiv:1103.1463].
\bibitem{cudell0}  K.~R.~Dienes and J.~-R.~Cudell, Phys.\ Rev.\ Lett.\  {\bf 72}, 187 (1994) [hep-th/9309126].
\bibitem{bronio} W.~Broniowski, W.~Florkowski and L.~Y.~Glozman, Phys.\ Rev.\ D {\bf 70}, 117503 (2004) [hep-ph/0407290].
\bibitem{TcTh0} B.~Lucini, M.~Teper and U.~Wenger, JHEP {\bf 0502}, 033 (2005) [hep-lat/0502003].
\bibitem{TcTh} B.~Bringoltz and M.~Teper, Phys.\ Rev.\  D {\bf 73}, 014517 (2006) [hep-lat/0508021].
\bibitem{zwie} B. Zwiebach (2004),\textit{ A First Course in String Theory} (Cambridge University Press, second edition, 2009).
\bibitem{cudell} T.~D.~Cohen, Phys.\ Lett.\ B {\bf 637}, 81 (2006) [hep-th/0602037].
\bibitem{cohen} T.~D.~Cohen, JHEP {\bf 1006}, 098 (2010) [arXiv:0901.0494].
\bibitem{meyer} H.~B.~Meyer, Phys.\ Rev.\ D {\bf 80}, 051502 (2009) [arXiv:0905.4229].
\bibitem{sve} B.~Svetitsky and L.~G.~Yaffe, Nucl.\ Phys.\ B {\bf 210}, 423 (1982).
\bibitem{Z3} R.~D.~Pisarski, Phys.\ Rev.\ D {\bf 62}, 111501 (2000) [hep-ph/0006205]; C.~Ratti, M.~A.~Thaler and W.~Weise, Phys.\ Rev.\ D {\bf 73}, 014019 (2006) [hep-ph/0506234].
\bibitem{G2} M.~Pepe and U.~-J.~Wiese, Nucl.\ Phys.\ B {\bf 768}, 21 (2007) [hep-lat/0610076].
\bibitem{diakonov} D. Diakonov and V. Petrov, arXiv:1011.5636, and references therein.
\bibitem{maas}  A.~Maas, JHEP {\bf 1102}, 076 (2011) [arXiv:1012.4284].
\bibitem{maas2}  V.~Macher, A.~Maas and R.~Alkofer, arXiv:1106.5381.
\bibitem{dashen} R. Dashen, S.-K. Ma and H.J. Bernstein, Phys. Rev. \textbf{187}, 345 (1969).
\bibitem{langelage10}  J.~Langelage and O.~Philipsen, JHEP {\bf 1004}, 055 (2010) [arXiv:1002.1507].
\bibitem{witten} E.~Witten, Nucl.\ Phys.\ B {\bf 160}, 57 (1979).
\bibitem{isgur} N.~Isgur and J.~E.~Paton, Phys.\ Rev.\ D {\bf 31}, 2910 (1985).
\bibitem{gluphen} F.~Buisseret, V.~Mathieu and C.~Semay, Phys.\ Rev.\ D {\bf 80}, 074021 (2009) [arXiv:0906.3098].
\bibitem{casi} L.~Del Debbio, H.~Panagopoulos, P.~Rossi and E.~Vicari, JHEP {\bf 0201}, 009 (2002) [hep-th/0111090].
\bibitem{casi2} A.~I.~Shoshi, F.~D.~Steffen, H.~G.~Dosch and H.~J.~Pirner, Phys.\ Rev.\ D {\bf 68}, 074004 (2003) [hep-ph/0211287].
\bibitem{g2r} W. E. Caswell, Phys. Rev. Lett. \textbf{33}, 244 (1974).
\bibitem{buiss11}  F.~Buisseret, Eur.\ Phys.\ J.\  C {\bf 71}, 1651 (2011) [arXiv:1101.0907].
\bibitem{glueb1} C.~J.~Morningstar and M.~J.~Peardon, Phys.\ Rev.\ D {\bf 60}, 034509 (1999) [hep-lat/9901004].
\bibitem{luciN} B.~Lucini, A.~Rago and E.~Rinaldi, JHEP {\bf 1008}, 119 (2010) [arXiv:1007.3879].
\bibitem{west} G.~B.~West, Phys.\ Rev.\ Lett.\  {\bf 77}, 2622 (1996) [hep-ph/9603316].
\bibitem{gershun10}  V.~D.~Gershun and A.~I.~Pashnev, Theor.\ Math.\ Phys.\  {\bf 73}, 1227 (1987); V.~D.~Gershun and D.~J.~Cirilo-Lombardo, J.\ Phys.\ A {\bf 43}, 305401 (2010).
\bibitem{borsa} Sz. Borsanyi, G. Endrodi, Z. Fodor, S.D. Katz and K.K. Szabo, arXiv:1104.0013.
\bibitem{Megias} E.~Megias, E.~Ruiz Arriola and L.~L.~Salcedo, Phys.\ Rev.\  D {\bf 80}, 056005 (2009)
  [arXiv:0903.1060]. 
\bibitem{case}  M.~Caselle, L.~Castagnini, A.~Feo, F.~Gliozzi and M.~Panero, JHEP {\bf 1106}, 142 (2011) [arXiv:1105.0359].
\bibitem{frautschi} S.~C.~Frautschi, Phys.\ Rev.\  D {\bf 3}, 2821 (1971).
\bibitem{cher} A.~Cherman, M.~Hanada and D.~Robles-Llana, Phys.\ Rev.\ Lett.\  {\bf 106}, 091603 (2011) [arXiv:1009.1623].
\bibitem{strange} F.~Buisseret, Eur.\ Phys.\ J.\ C {\bf 68}, 473 (2010) [arXiv:0912.0678].
\bibitem{highT} G. Endrodi, Z. Fodor, S.D. Katz and K.K. Szabo, PoS {\bf LATTICE2007}, 228 (2007) [arXiv:0710.4197].
\bibitem{Laine} K.~Kajantie, M.~Laine, K.~Rummukainen and Y.~Schroder, Phys.\ Rev.\  D {\bf 67}, 105008 (2003)
  [hep-ph/0211321].
\bibitem{Laine2} F.~Di Renzo, M.~Laine, V.~Miccio, Y.~Schroder and C.~Torrero, JHEP {\bf 0607}, 026 (2006)
  [hep-ph/0605042].
\bibitem{HTL} J.~O.~Andersen, M.~Strickland and N.~Su, Phys.\ Rev.\ Lett.\  {\bf 104}, 122003 (2010) [arXiv:0911.0676]; JHEP {\bf 1008}, 113 (2010) [arXiv:1005.1603].
\bibitem{panero} M. Panero, Phys. Rev. Lett. \textbf{103}, 232001 (2009) [arXiv:0907.3719].
\bibitem{gupta} S.~Datta and S.~Gupta, Phys.\ Rev.\  D {\bf 82}, 114505 (2010) [arXiv:1006.0938].
\bibitem{scalT} S. Gupta, K. Huebner and O. Kaczmarek, Phys. Rev. D {\bf 77}, 034503 (2008) [arXiv:0711.2251].
\bibitem{brau} F.~Brau and F.~Buisseret, Phys.\ Rev.\ D {\bf 79}, 114007 (2009) [arXiv:0902.4836].
\bibitem{buiss10} F.~Buisseret and G.~Lacroix, Eur.\ Phys.\ J.\  C {\bf 70}, 1051 (2010) [arXiv:1006.0655].
\bibitem{holl} K.~Holland, M.~Pepe and U.~J.~Wiese, Nucl.\ Phys.\ B {\bf 694}, 35 (2004) [hep-lat/0312022].
\bibitem{braun} J.~Braun, A.~Eichhorn, H.~Gies and J.~M.~Pawlowski, Eur.\ Phys.\ J.\ C {\bf 70}, 689 (2010)  [arXiv:1007.2619].
\end{thebibliography}
\end{document}